\documentclass[aps, prd, nofootinbib, twocolumn, preprintnumbers, showpacs, floatfix]{revtex4}
\usepackage{graphicx}
\usepackage{amsmath}
\usepackage{multirow}
\usepackage{bm}
\usepackage{color}
\def\fsl#1{\setbox0=\hbox{$#1$}                 
   \dimen0=\wd0                                 
   \setbox1=\hbox{/} \dimen1=\wd1               
   \ifdim\dimen0>\dimen1                        
      \rlap{\hbox to \dimen0{\hfil/\hfil}}      
      #1                                        
   \else                                        
      \rlap{\hbox to \dimen1{\hfil$#1$\hfil}}   
      /                                         
      \fi}                                      %

\begin{document}
\title{A note on the coupling of the techni-dilaton to the weak bosons}
\author{Michio Hashimoto}
 \email{michioh@isc.chubu.ac.jp}
  \affiliation{
   Chubu University, \\
   1200 Matsumoto-cho, Kasugai-shi, \\
   Aichi, 487-8501, JAPAN}
\date{\today}

\begin{abstract}
In this note, we study the coupling of the techni-dilaton 
to the weak bosons.
We consider two cases:
(1) The dilaton directly couples to the weak bosons 
similarly to the SM.
(2) The coupling in question is effectively induced only through
the techni-fermion loops.
In both cases, we find that the coupling is essentially determined by
the mass-squared of the weak bosons over the dilaton decay constant.
\end{abstract}

\maketitle

One of the most important aim at the Tevatron and 
at the CERN Large Hadron Collider (LHC) is to discover the Higgs boson.
The direct searches of the standard model (SM) Higgs boson at the LEP
have set limits on the Higgs mass to be larger than 114.4~GeV~\cite{pdg}.
Recently, the mass ranges of the SM Higgs boson from 114~GeV to 600~GeV
have been narrowed down to several windows and 
slits~\cite{ATLAS-Higgs-search,CMS-Higgs-search,CDF:2011cb}.
The fourth generation model~\cite{He:2001tp,Hashimoto:2009xi,Frampton:1999xi} 
are also constrained~\cite{ATLAS-Higgs-search,CMS-4G-search}.
Besides, these results impact on several classes of 
the top condensate models~\cite{Hill:2002ap}.

A heavy Higgs boson can be a signal of 
the existence of models beyond the SM (BSM), 
because non-standard contributions
to the $S$ and $T$ parameters~\cite{Peskin-Takeuchi} are required 
for consistency with the LEP precision measurements~\cite{pdg}.
Such a class of the models contains the walking technicolor (WTC)
scenario~\cite{Holdom:1981rm,Yamawaki:1985zg,Akiba:1985rr,Appelquist86}.

It is believed that in the WTC, there appears a scalar particle, 
so-called the techni-dilaton (TD),
which is the pseudo Nambu-Goldstone (NG) boson associated with 
the scale symmetry breaking~\cite{Yamawaki:1985zg,Bando:1986bg}.
The TD mass near the critical point has been suggested as 
$M_{\rm TD} \sim \sqrt{2}m$ in the context of 
the gauged Nambu-Jona-Lasinio (NJL) model~\cite{Shuto:1989te}, 
where $m$ represents the dynamically generated fermion mass.
The TD mass in the criticality limit is discussed recently in
Refs.~\cite{Hashimoto:2010nw,Dietrich:2005jn}.

In the previous work~\cite{Hashimoto:2011ma}, 
we have studied the yukawa couplings of the SM fermions in the WTC,
because the gluon fusion process, which is important in 
the heavy Higgs searches, depends on the magnitude of the yukawa coupling
in addition to the trivial factor arising from 
the number of the extra heavy colored particles.

In this note, we briefly analyze the coupling of the TD to the weak bosons.

Let us first consider the case that 
the dilaton $\sigma$ directly couples to $W$.
This situation is similar to the SM.

\begin{figure}[t]
   \begin{center}
   \resizebox{0.4\textwidth}{!}{\includegraphics{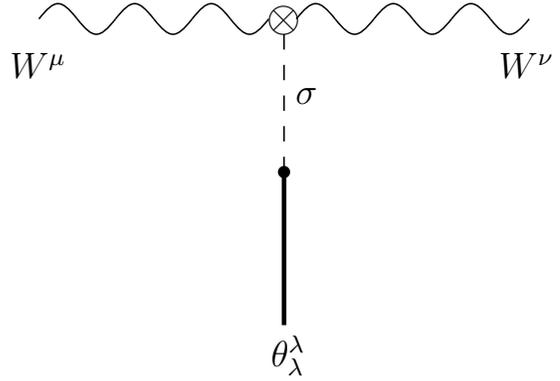}}
   \end{center}
   \caption{The $\sigma WW$ coupling in the case 
    that the $\sigma$ directly couples to $W$.
   \label{fig-sWW}}
\end{figure}

Owing to the nature of the energy-momentum tensor,
we formally obtain the following relation~\cite{donoghue}, 
\begin{equation}
  \langle W(p)| \theta_\lambda^\lambda (0) | W (p) \rangle = 2M_W^2\,.
  \label{amp-sWW}
\end{equation}
Assuming the $\sigma$ dominance at the zero momentum transfer
as shown in Fig.~\ref{fig-sWW},
we can read the $g^{\mu\nu}$-part of 
the $\sigma$--$W^\mu$--$W^\nu$ form factor $\Gamma_{\sigma WW}^{\mu\nu}$ as 
\begin{equation}
  g_{\sigma WW}(0) = \frac{2M_W^2}{F_\sigma}, \label{sWW}
\end{equation}
where $F_\sigma$ represents the dilaton decay constant being
$\langle 0| \theta_\lambda^\lambda (0)|\sigma (q) \rangle = F_\sigma M_\sigma^2$
with the dilaton mass $M_\sigma$.
The expression (\ref{sWW}) agrees with 
the result in Refs.~\cite{Goldberger:2007zk,Fan:2008jk}. 
For a generalization of Refs.~\cite{Goldberger:2007zk,Fan:2008jk}, 
see also Ref.~\cite{Vecchi:2010gj}.

Next, we study, so-called, the techni-dilaton $\sigma_{T}^{}$ which 
couples to $W$ only through the techni-fermions (TF's).
The axial current $J_A^\mu$ of the TF's yields
the decay constant $F_\pi$, 
$\langle 0| J_A^\mu(0)|\pi (q) \rangle = -iq^\mu F_\pi$,
and the weak boson mass is provided by $F_\pi$. 
We thus consider the coupling between $\sigma_T^{}$ and $J_A^\mu$.

\begin{figure}[t]
   \begin{center}
   \resizebox{0.4\textwidth}{!}{\includegraphics{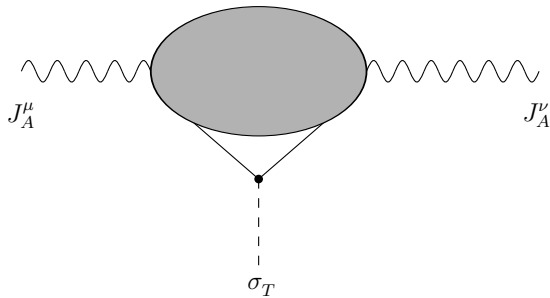}}
   \end{center}
   \caption{Coupling of the TD to the axial currents of the TF's.
   The TD $\sigma_T^{}$ couples to $J_A^\mu$ only through 
   the internal TF lines.
   \label{fig-sAA}}
\end{figure}

The axial current correlator in the momentum space is 
\begin{equation}
  {\rm F.T.}i\langle 0| J_A^\mu(x) J_A^\nu(0)|0\rangle = 
  \left(g^{\mu\nu}-\frac{q^\mu q^\nu}{q^2}\right)\Pi_A(q^2) \, .
\end{equation}
The vacuum polarization function $\Pi_A$ is characterized by
\begin{equation}
  \Pi_A (0) = F_\pi^2 \, .
\end{equation}
This relation plays an important role in our approach.

The $\sigma_T^{}$ coupling to $J_A^\mu$ at the zero momentum transfer
is just like the mass insertion:
Note that the identity holds
\begin{equation}
  \frac{1}{\fsl{\ell} - m} y_T^{} \frac{1}{\fsl{\ell} - m} 
  = y_T^{} \frac{\partial}{\partial m} \frac{1}{\fsl{\ell} - m} ,
\end{equation}
where $m$ and $y_T^{}$ are the dynamically generated TF mass and 
the yukawa coupling, respectively.
We can then obtain the coupling of $\sigma_T^{}$ to $J_A^\mu$
at zero momentum simply by
\begin{equation}
  g_{\sigma_T^{} AA}(0) = y_T^{} \frac{\partial \Pi_A(0)}{\partial m}\,.
  \label{sT-AA} 
\end{equation}
Because $F_\pi$ is generated through the TF loop effects,
$F_\pi$ should be proportional to $m$, i.e., $F_\pi = \kappa \, m$,
when we take the infinite limit of the ETC scale. 
Even in a realistic situation with a finite ETC scale 
$\sim {\cal O}(1000 \mbox{ TeV})$,
we expect that $F_\pi$ does not strongly depend on the ETC scale.
One could find the numerical factor $\kappa$ in 
Ref.~\cite{Hashimoto:2011ma}, 
$\kappa \equiv \kappa_F \sqrt{N_{\rm TC}}/(2\pi)$ with
$\kappa_F \simeq 1.4\mbox{--}1.5$ and 
$N_{\rm TC}$ being the number of the color of the TC gauge group, 
where the Pagels-Stokar formula~\cite{Pagels:1979hd} is employed. 
Then Eq.~(\ref{sT-AA}) yields
\begin{equation}
  g_{\sigma_T^{} AA}(0) = y_T^{} \frac{2F_\pi^2}{m} \, .
\end{equation}
Attaching $W^\mu$ to $J_A^{\mu}$,
we finally obtain the coupling of the TD to the weak bosons
at zero momentum,
\begin{equation}
  g_{\sigma_T^{} WW}(0) = y_T^{} \frac{2M_W^2}{m} \, .
  \label{sT-WW}
\end{equation}

The two cases are conceptually different.
However, when the yukawa coupling is like the SM, $y_T^{} = m/F_\sigma$,
Eq.~(\ref{sT-WW}) formally agrees with Eq.~(\ref{sWW}).
The yukawa coupling was also estimated as $y_T^{} = (3-\gamma_m)m/F_\sigma$
with the anomalous dimension $\gamma_m \; (\simeq 1)$
for the model in Ref.~\cite{Bando:1986bg},
where the four-fermion interactions were incorporated,
${\cal L} = {\cal L}_{\rm TC} + G_1 (\bar{T}T)^2 + G_2 (\bar{T}T) (\bar{f}f)
 + G_3 (\bar{f}f)^2$ with ${\cal L}_{\rm TC}$ standing for 
the TC gauge theory, and $T$ and $f$ being the TF's and 
the SM fermions, respectively. 
If so, this suggests that $g_{\sigma_T^{} WW}$ is changed by 
the additional factor $(3-\gamma_m)$ from Eq.~(\ref{sWW}).
Therefore we conclude that the coupling of the TD to the weak bosons is
essentially determined by the mass-squared of the weak bosons
over the TD decay constant.

Although we have estimated the coupling $g_{\sigma_{(T)}^{} WW}$ at zero momentum,
one might expect that 
the on-shell one is not so far from these estimates.
Strictly speaking, the TF mass function in the internal line
is not a constant $m$.
In sufficiently low energy, however, this would not affect
the estimate so much.

The results derived in this note mean that 
the (effectively induced) operator
$\frac{\sigma_T^{}}{F_\sigma} W_\mu W^\mu$ yields 
the coupling between the TD and the weak bosons,
similarly to the SM.
The earlier argument in Ref.~\cite{Matsuzaki:2011qt}
contradicted ours, i.e.,
they argued that the higher dimensional operator 
$\frac{\sigma_T^{}}{F_\sigma}W_{\mu\nu}W^{\mu\nu}$ gave
the $\sigma_T^{}$--$W$--$W$ coupling 
when the TD couples to $W$ only through the TF-loop.
In the end they have revised it, 
following our results~\cite{Matsuzaki:2011ie}. 

In any case, the Higgs boson might be revealed soon.
What exciting data will be supplied at the LHC and the Tevatron?

\end{document}